\pdfoutput=1
\documentclass[twocolumn]{aastex62}
\usepackage{graphicx}

\submitjournal{ApJ}

\shorttitle{Multiphase gas flows in ESO428$-$G14}
\shortauthors{Feruglio et al.}

\begin{document}

\title{MULTIPHASE GAS FLOWS IN THE NEARBY SEYFERT GALAXY ESO428-G14.}

\correspondingauthor{Chiara Feruglio}
\email{chiara.feruglio@inaf.it}

\author[0000-0002-4227-6035]{C. Feruglio}
\affil{INAF Osservatorio Astronomico di Trieste, 
Via G.B. Tiepolo, 11, Trieste, I-34143, Italy}

\author[0000-0002-3554-3318]{G. Fabbiano}
\affil{Center for Astrophysics, Harvard \& Smithsonian, 
60 Garden St. Cambridge, MA 02138, USA}

\author[0000-0002-4314-021X]{M. Bischetti}
\affil{INAF Osservatorio Astronomico di Roma, Via Frascati 33, 00078 Monteporzio Catone, Italy}

\author{M. Elvis}
\affil{Center for Astrophysics, Harvard \& Smithsonian, 
60 Garden St. Cambridge, MA 02138, USA}

\author{A. Travascio}
\affil{INAF Osservatorio Astronomico di Roma, Via Frascati 33, 00078 Monteporzio Catone, Italy}

\author{F. Fiore}
\affil{INAF Osservatorio Astronomico di Trieste, 
Via G.B. Tiepolo, 11, Trieste, I-34143, Italy}

\begin{abstract}

We present ALMA rest-frame 230 GHz continuum and CO(2-1) line observations of the nearby Compton-thick Seyfert galaxy ESO428-G14, with angular resolution 0.7 arcsec (78 pc).  
We detect CO(2-1) emission from spiral arms and a circum-nuclear ring with 200 pc radius, and 
from a transverse gas lane with size of $\sim100$ pc, which crosses the nucleus and connects the two portions the circumnuclear ring. 
The molecular gas in the host galaxy is distributed in a rotating disk with intrinsic circular velocity 
$v_{rot}=135$ km/s, inclination $i=57$ deg, and dynamical mass $M_{dyn }=5\times 10^9~\rm M_{\odot}$ within a radius of $\sim 1$ kpc.
In the inner 100 pc region CO is distributed in a equatorial bar, whose kinematics is highly perturbed and consistent with an inflow of gas towards the AGN. 
This inner CO bar overlaps with the most obscured, Compton-thick region seen in X-rays. We derive a column density of $\rm N(H_2) \approx 2\times10^{23}~ cm^{-2}$ in this region, suggesting that molecular gas may contribute significantly to the AGN obscuration. 
We detect  a molecular outflow with a total outflow rate $\rm \dot M_{of}\approx 0.8~M_{\odot}/yr$, distributed along a bi-conical structure with size of $700$ pc on both sides of the AGN. The bi-conical outflow is also detected in the $\rm H_2$ emission line at 2.12 $\mu$m, which traces a warmer nuclear outflow located within 170 pc from the AGN. 
This suggests that the outflow cools with increasing distance from the AGN. 
We find that the hard X-ray emitting nuclear region mapped with Chandra is CO-deprived, but filled with warm molecular gas traced by $\rm H_2$ - thus confirming that the hard (3-6 keV) continuum and Fe K$\alpha$ emission are due to scattering from dense neutral clouds in the ISM.

\end{abstract}

\keywords{
galaxies: active -- galaxies: ISM -- galaxies: kinematics and dynamics -- galaxies: Seyfert -- X-rays: galaxies
}

\section{Introduction}

The nearby  Seyfert galaxy ESO428-G14 (D$\sim$23.3 Mpc, NED; which corresponds to a physical scale of 112 pc/arcsec) hosts a Compton-thick (CT) AGN (Maiolino et al 1998; Risaliti et al 1999), with intrinsic bolometric luminosity $\rm L_{bol, AGN}=4.1\times 10^{43}~erg/s$ (Levenson et al. 2006).  
According to Fabbiano et al. (2019), the likely black hole mass in ESO 428-G014 is $(1-3) \times 10^7 M_{\odot}$, using the McConnell \& Ma (2013) scaling from bulge velocity dispersion, $\sigma  =119.7$ km/s, and bulge mass, $\rm M(K)_{bulge} = 10^{9.14}~ M_{\odot}$ (Peng et al. 2006). 
The latter is inferred from the K band magnitude and assuming a mass-to-light ratio of 0.6 (McGaugh \& Schombert 2014). 
Given this black hole mass estimate, the Eddington ratio of ESO 428-G014 is about 1\%.
The host galaxy of ESO428-G14 has a stellar mass of $\rm M_*=3.76\times 10^{10}~M_{\odot}$ and a star-formation rate of $\rm SFR_{FUV}=1.5\times 10^{-4}~M_{\odot}/yr$ estimated from the FUV luminosity (Vaddi et al. 2016). The latter is a lower limit SFR because it is not corrected for extinction at the rest frame. 
ESO4280-G14 is known to have a radio jet (Falcke et al. 1996 and references therein), which, according to Riffel et al. (2007) is launched at a small inclination with respect to the plane of the disk, such that it impacts the disk and it entrains [OIII] emitting gas from the disk plane on scales of 300-600 pc in the Narrow Line Region of the AGN (Falcke et al. 1996). Ionized gas, probably pushed by the radio jet, is detected in several ionized gas tracers, such as  [OIII] and H$\alpha$ with HST and Gemini (Falcke et al. 1996, Riffel et al. 2007), and warm $H_2$ with SINFONI/VLT (May et al. 2018). 

ESO428-G014 has been the subject of four recent papers, based on deep Chandra ACIS observations. Fabbiano et al. (2017, 2018a; Papers I and II) detected kiloparsec-size extended regions emitting in hard X-ray (3-6 keV) continuum and Fe K$\alpha$ line.  Chandra high resolution imaging of the soft ($<$3 keV), line-dominated X-ray emission (Fabbiano et al 2018b, Paper III) shows remarkable correspondence between high surface brightness X-ray features and the radio jet and optical line emission. Paper III also shows that in the inner circum-nuclear region of $\sim150$ parsec, the X-ray hardness ratios are larger, indicating obscuration by Compton thick material. In this inner region, higher surface brightness extended emission is present in both hard ($>3$ keV) continuum and 6.4 keV Iron emission line (Fabbiano et. al 2019, Paper IV). 

Paper IV also reports clumpy features in the Iron line emission of ESO428-G014. Questions arise about the nature of these circum-nuclear hard X-ray features, whether, e.g., they are due to a nuclear disk of dense material (the torus), or to the interaction of hard AGN photons with dense molecular clouds in the nuclear outflow, responsible for the ionization cone emission. In a similar CT AGN, NGC5643, also recently imaged with Chandra, a distinct X-ray feature in the 6.4 keV Iron line was reported (Fabbiano et al 2018c), corresponding with the rotating nuclear disk found with ALMA by Alonso-Herrero et al. (2018).

All these extended and complex hard X-ray features, both on kiloparsec and $\sim100$ pc scale, require dense scattering clouds in the interstellar medium. 
ALMA observations of molecular gas tracers with high enough angular resolution have the potential to shed light on the cold gas and dust component in the host galaxy and in the inner nuclear region. To this aim we have checked the ALMA archive and found public data of the rest-frame 230 GHz continuum and CO(2-1) line observations of ESO428-G14, with an angular resolution of 0.7 arcsec (78 pc).

The paper is organized as follows. Section 2 presents the ALMA observational setup and data analysis. Section 3 presents observational results. In Section 4 we compare the emission from cold molecular gas and dust seen by ALMA, with the X-ray emission seen by Chandra in different energy bands, and discuss our results.

\begin{figure*}[t]
\centering
\includegraphics[width=14cm]{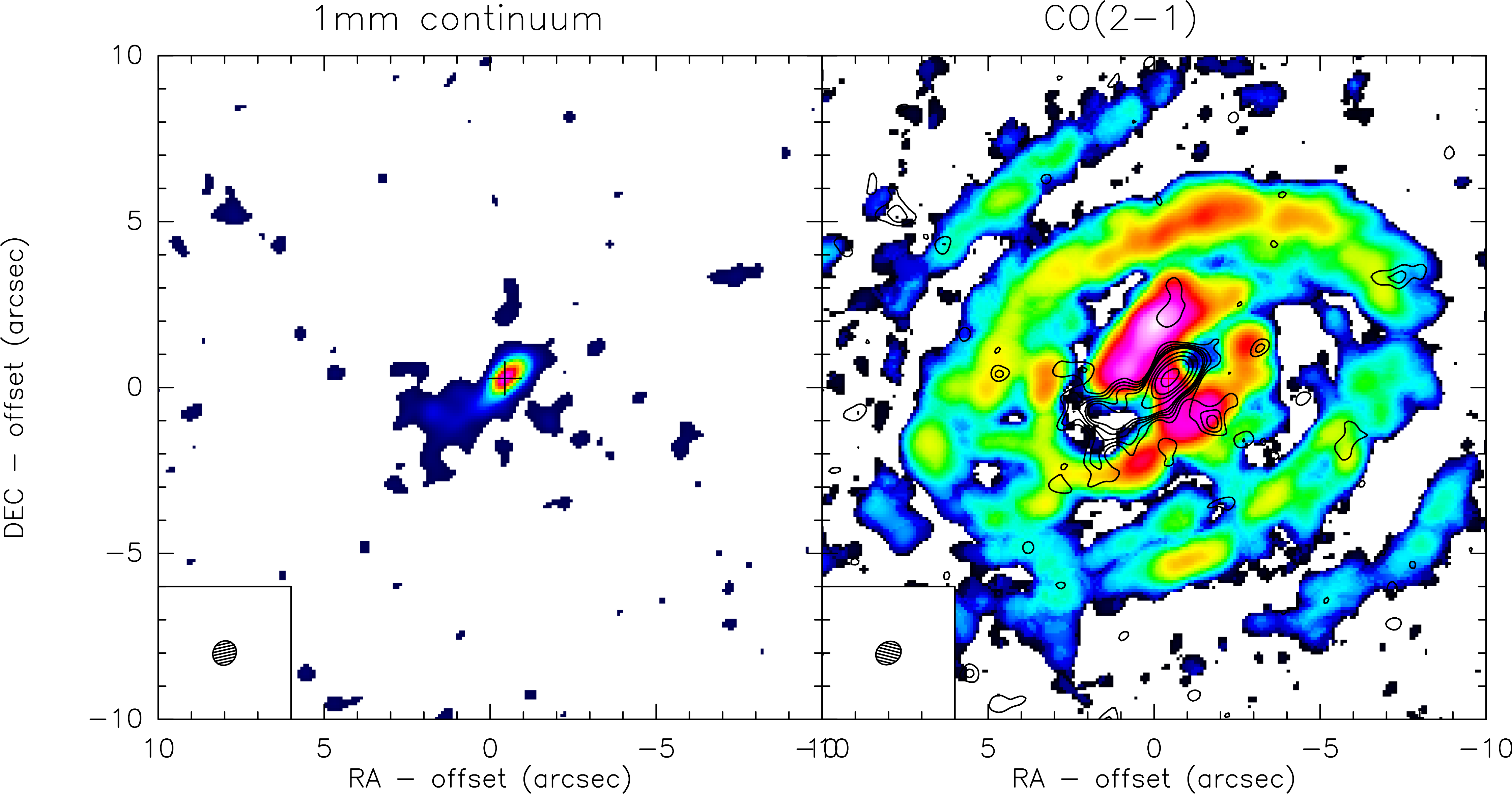}
\caption{Left panel: the 1.3 mm continuum map of ESO428-G14. Regions with emission below 2$\sigma$ have been blanked ($\sigma=0.027$ mJy/beam). Contours are drawn at $2,3,5,10,15,20\sigma$.
Right panel: the CO(2-1) mean-flux (moment 0) map . 
A detection threshold of 3.5$\sigma$ has been applied to produce the map. 
The black contours show the 1.3 mm continuum emission (same as in the left panel).
The synthesized beam, $0.8\times0.67$ arcsec FWHM at $\rm PA=-59$ deg, is shown in the insets. }
\end{figure*}

\begin{figure*}
\includegraphics[width=\textwidth]{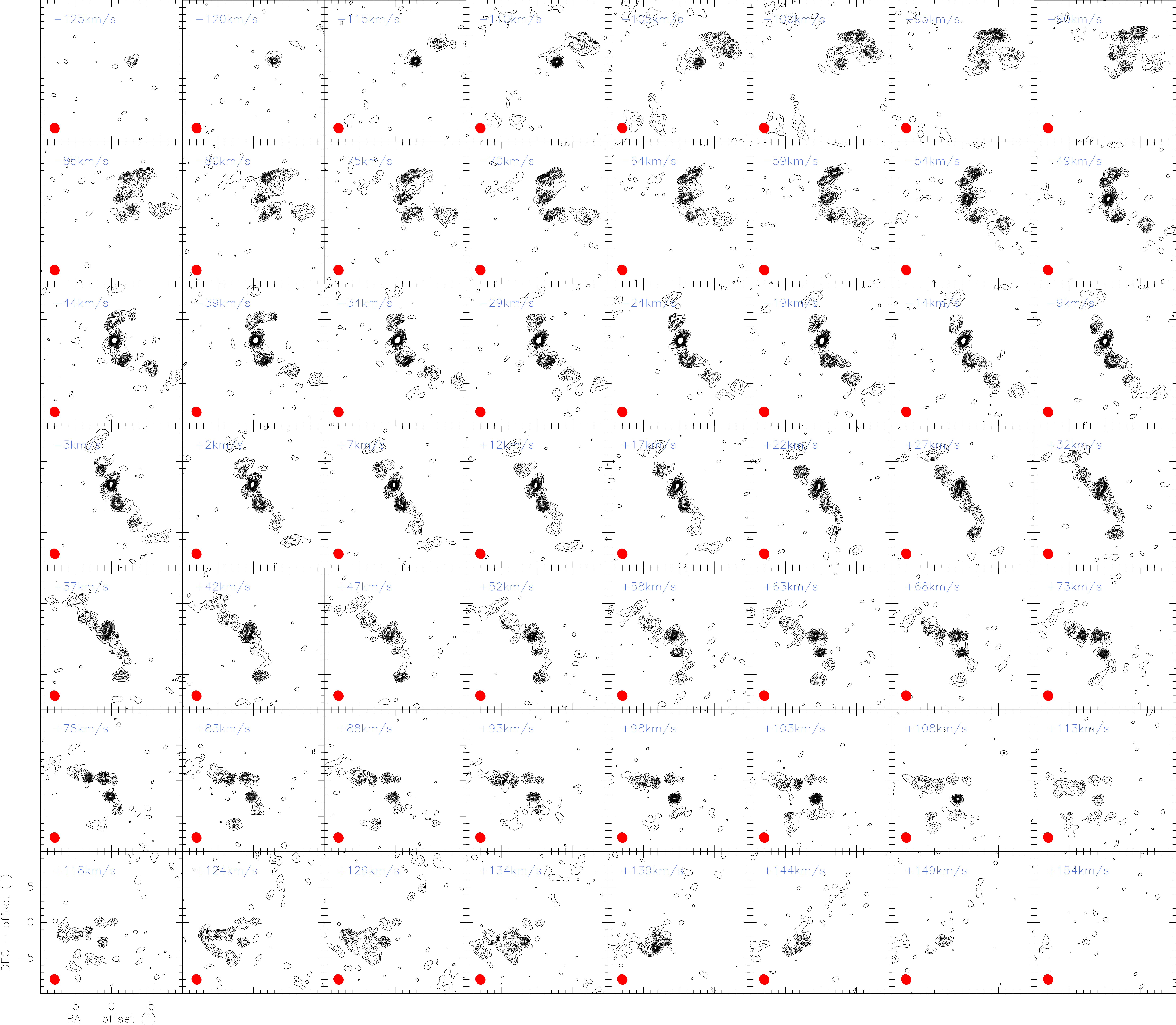}
 \caption{CO(2-1) emission channel maps in velocity bins of $\sim 5 $ km/s. Each box is 1.1 kpc on the side. The velocity of each map relative to the systemic one is indicated in the labels. The synthetic beam is show in the bottom-left corner by a red ellipse. Contours are 2$\sigma$ to 40$\sigma$ in steps of 2$\sigma$, $\sigma=1.16$ mJy/beam. 
 }
\end{figure*}

\begin{figure*}
\includegraphics[width=\textwidth]{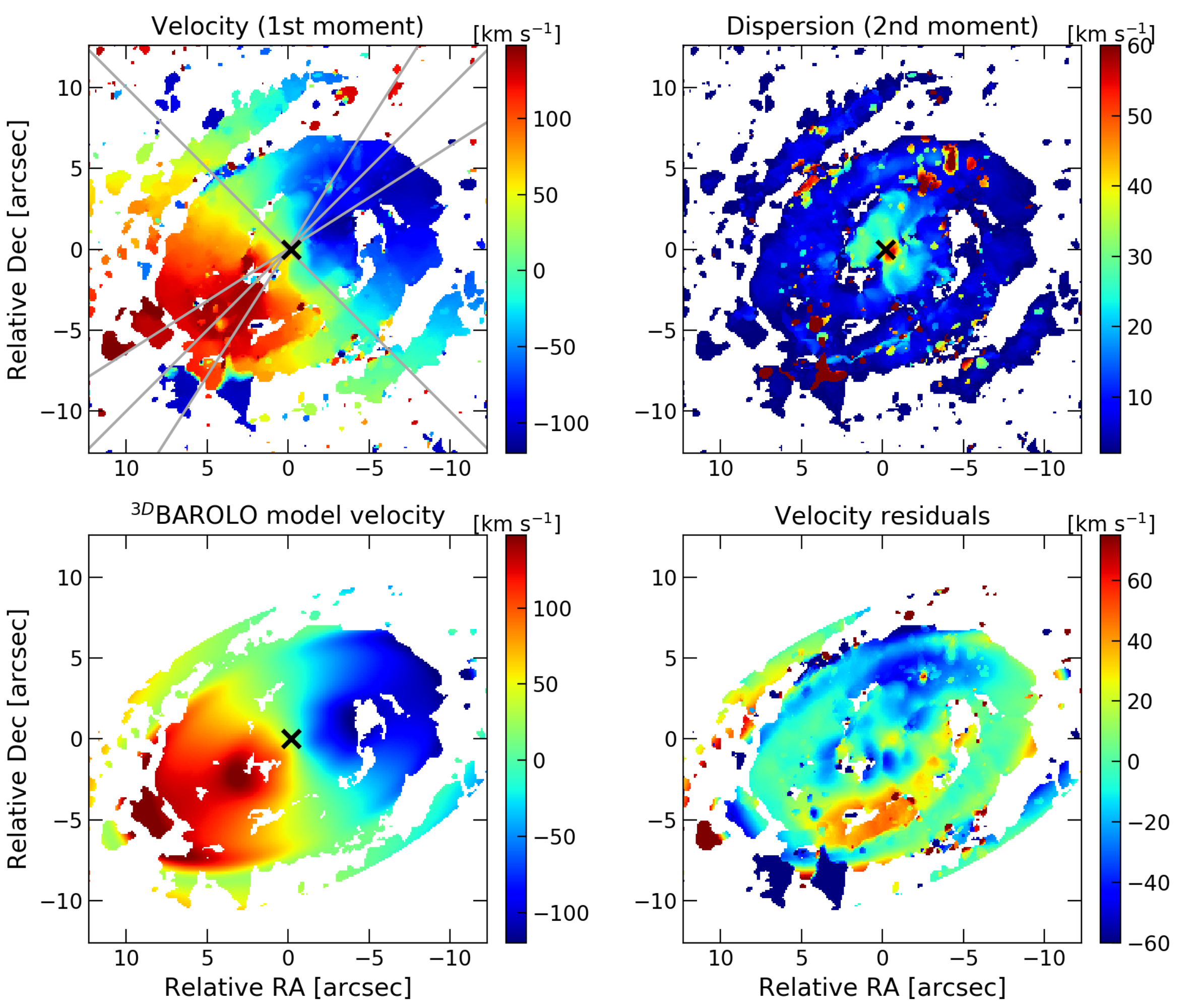}
\caption{The CO(2-1) mean-velocity (upper-left) and velocity dispersion maps (upper-right panel).  
A detection threshold of 3.5$\sigma$ was applied to derive the moment maps. The cross indicates the position of the 1.3 mm continuum peak. 
The grey lines indicate the orientation of the slices used to derive the PV diagrams in Fig. 5.
Bottom-left panel: the mean-velocity map of the $\rm ^{3D}BAROLO$ disk model. 
Bottom-right panel: the residual mean-velocity field map ($data-model$). 
Color bars are in units of km/s. 
}
\end{figure*}

\begin{figure}
\centering
\includegraphics[width=8cm]{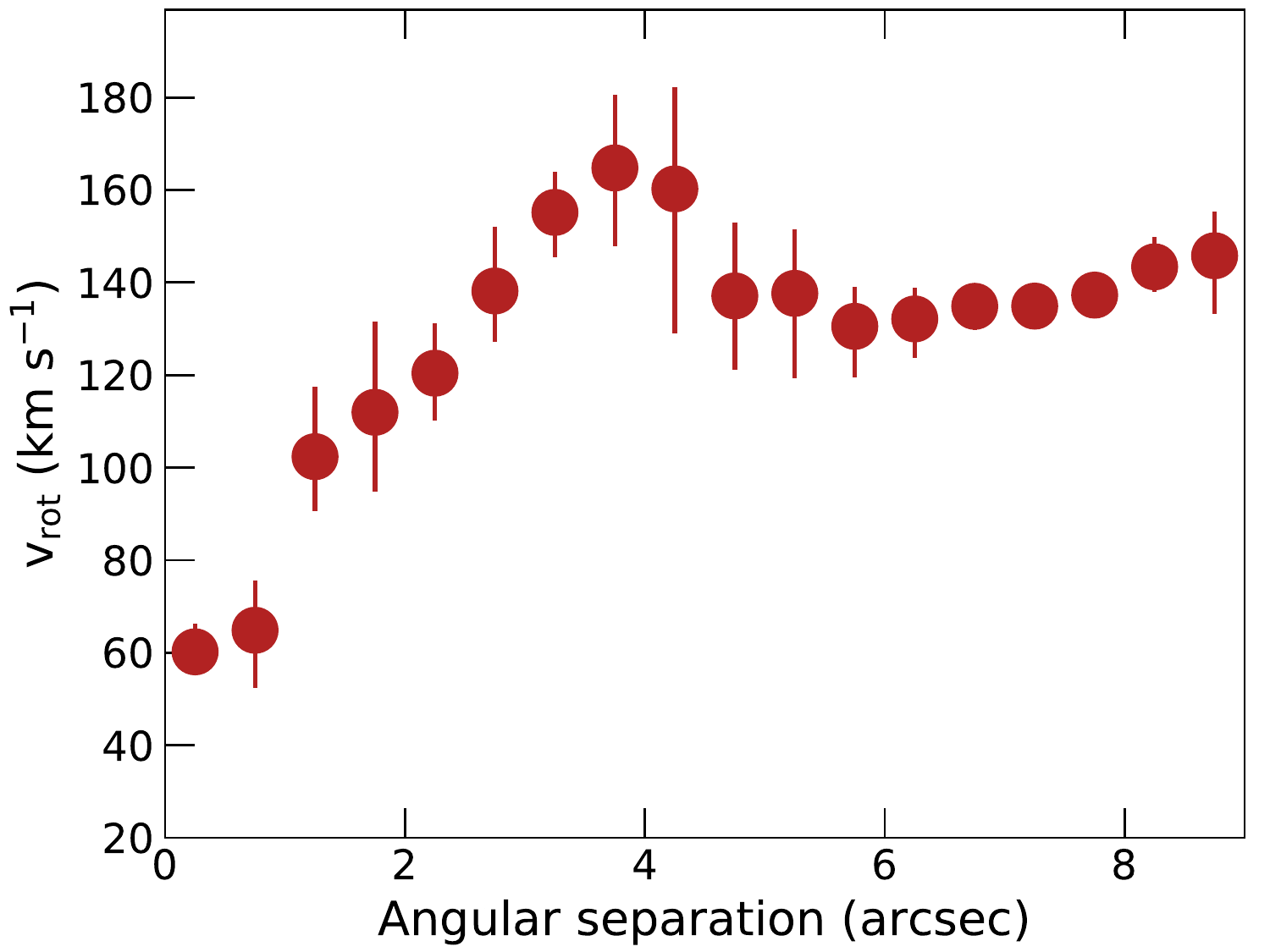}\\
\caption{The rotation velocity $\rm v_{rot}$ versus the radius of the best-fit inclined disk model derived by $\rm ^{3D}BAROLO$. $\rm v_{rot}$ increases from the centre, shows an peak between radii 2 and 4 arcsec, and then reaches a plateau at 135 km/s in the outer part of the disk.
}
\end{figure}

\section{Data and Analysis}
We retrieved ALMA and optical/NIR IFU VLT data from the respective archives, as described below. 
We also used archival Chandra data, as analyzed in Paper III.

\subsection{ALMA} 
We retrieved data of ESO428-G14 from the ALMA archive (program ID 2015.1.00086.S). 
These are band 6 observations acquired with the 12 m array during May 2016, in the frequency ranges 228-232 and 
242.6-246.6 GHz, which cover the CO(2-1) line and the underlying 1.3 mm continuum. 
The data have maximum  spectral resolution of 2.5 km/s. Observation time, including overheads, was 29 minutes. 
We created the calibrated measurement set under software CASA (McMullin et al. 2007). 
We then converted it into a visibility table in GILDAS format, and we imaged the data using Mapping (Guilloteau \& Lucas 2000).
To obtain the continuum map, we first averaged the visibilities combining  the four available spectral windows and excluding the range set by the CO(2-1) sky frequency plus/minus 300 km/s, to stay in the spectral region free from the emission line. We then imaged the continuum visibilities by adopting a natural weighting scheme, and produced a clean cube by using the \texttt{mx} cleaning algorithm down to a detection threshold 0.5 times the r.m.s. noise. 
This way we obtain a synthesised beam of $0.76\times0.69$ arcsec$^2$ at a $\rm PA= ˆ'32$ deg. 
The reached r.m.s. noise in the clean map is 0.027 mJy/beam in the continuum in the aggregated bandwidth. 
We also deconvolved using the \texttt{MRC} cleaning algorithm (Wakker \& Schwarz 1988) to highlight multi scale emission. We obtain consistent results between 
\texttt{mx} and  \texttt{MRC}. 
To obtain the continuum-subtracted visibilities of the CO(2-1) line, we subtracted the continuum in the {\it uv-plane} from the visibility set using the task {\it uv-subtract} within GILDAS. 
We then imaged the CO(2-1) visibilities using a natural weighting scheme with detection threshold 0.5 times the noise, reaching a r.m.s. noise level of 1.25 mJy/beam in 2.5 km s$^{-1}$ wide channels after cleaning, and a synthetic beam of $0.8\times0.67$ arcsec$^2$ FWHM at $\rm PA= -59$ deg. 

In order to better map the inner compact region, we produced also a CO(2-1) data cube with enhanced angular resolution.  
We used two techniques. 
First, we imaged and de-convolved applying Robust weighting within GILDAS, and obtained a synthetic beam of $0.57\times0.53$ arcsec FWHM at $PA=-55$ and a r.m.s. $5.8\times 10^{-4}$ Jy/beam per 2.5 km/s channel. 
The second method consists in further enhancing the angular resolution by restoring with a clean beam size of $0.4\times0.4$ arcsec, chosen to approximately match the {\it Chandra} PSF (super-resolution). We derived for this map a r.m.s. noise of $5.8\times 10^{-4}$ Jy/beam per 2.5 km/s channel. 
Super-resolution can break the flux estimation. For our case we checked that flux densities in the super-resolution map agrees within 15-20\%  with those of the Robust weighted images. In any case, we used the super-resolution map only for describing the morphology and kinematics of the inner compact region, but not for estimating fluxes and derived quantities

\begin{figure*}
\centering
\includegraphics[width=7.5cm]{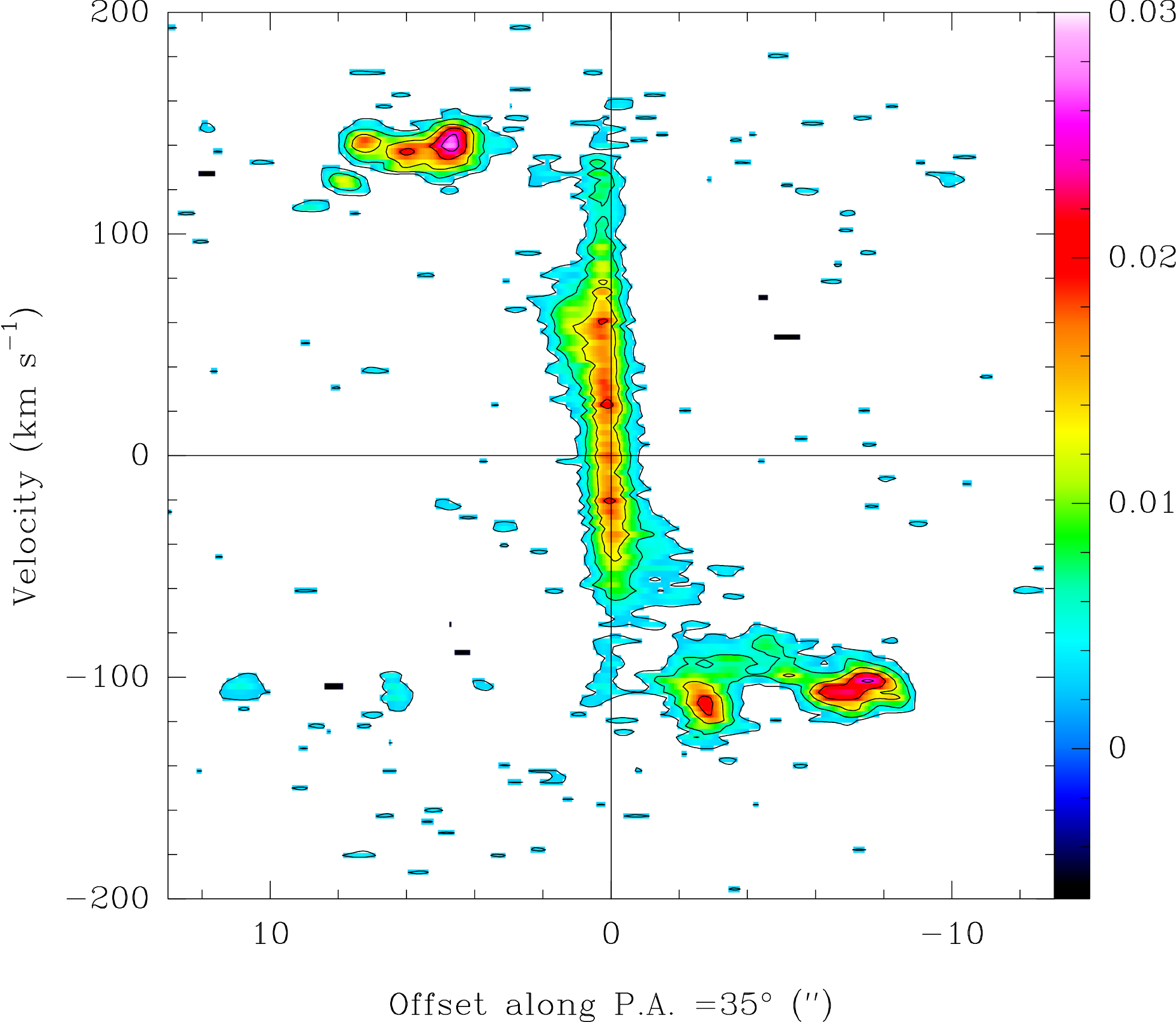}
\includegraphics[width=7.5cm]{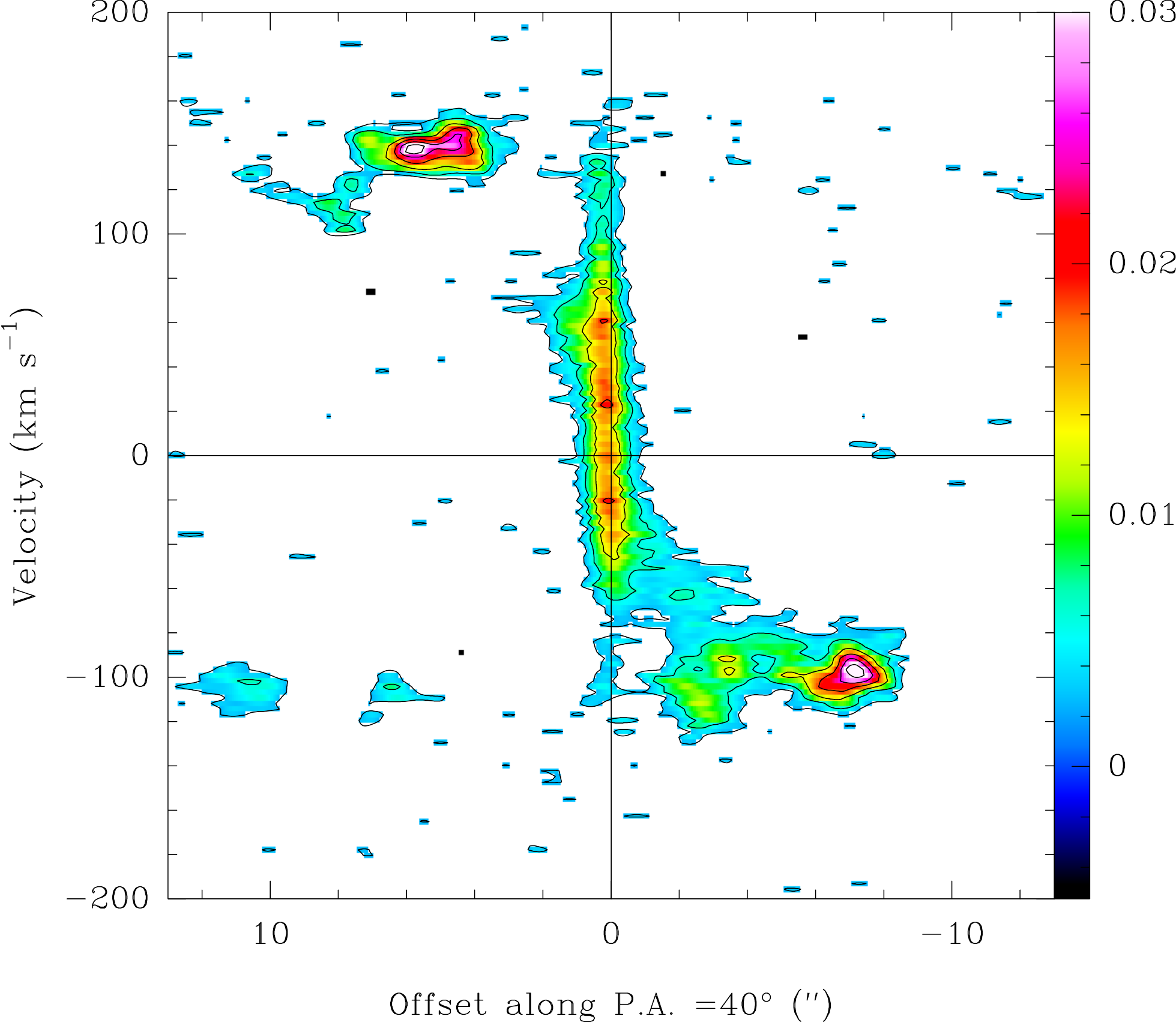}
\includegraphics[width=7.5cm]{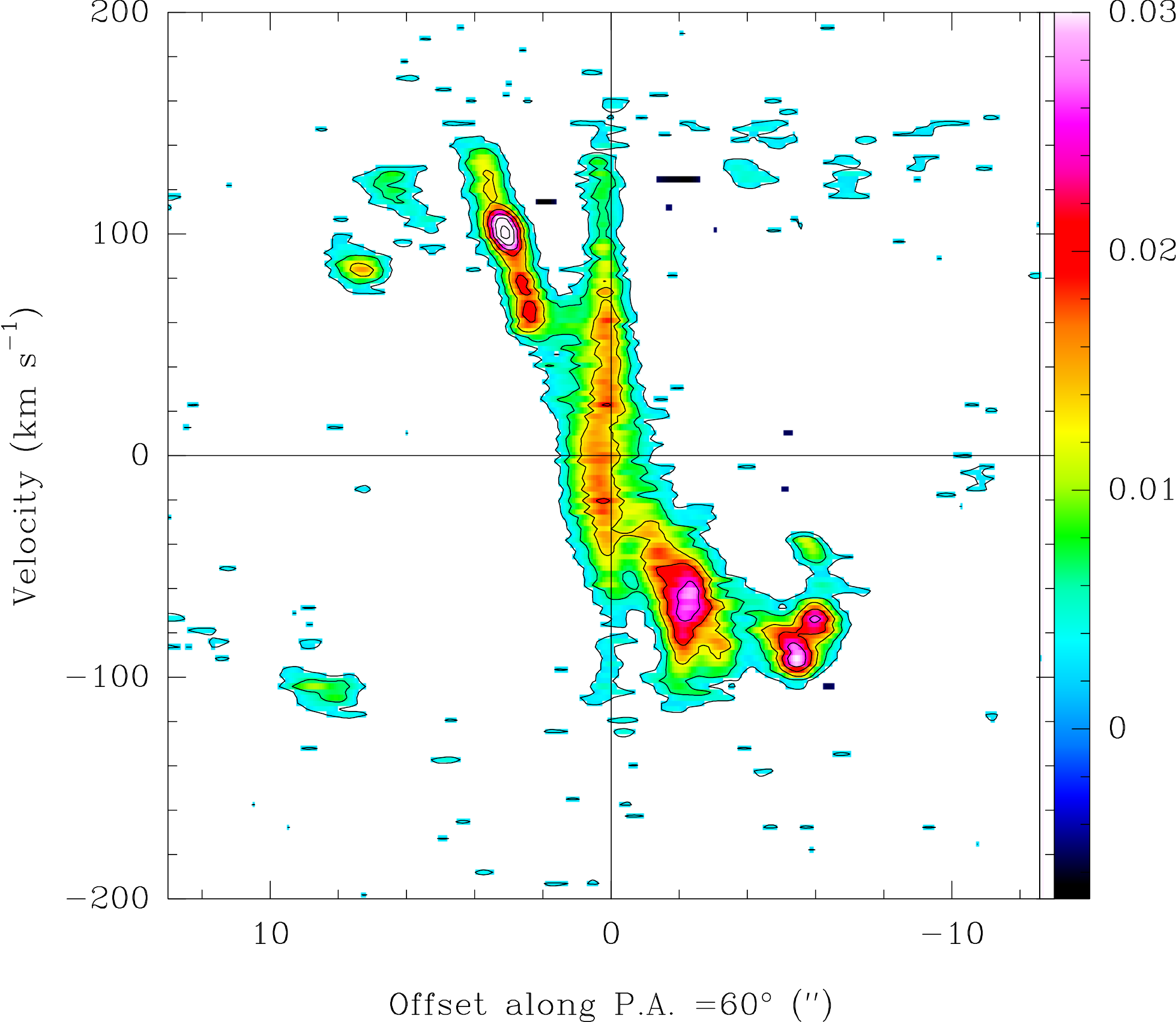}
\includegraphics[width=7.5cm]{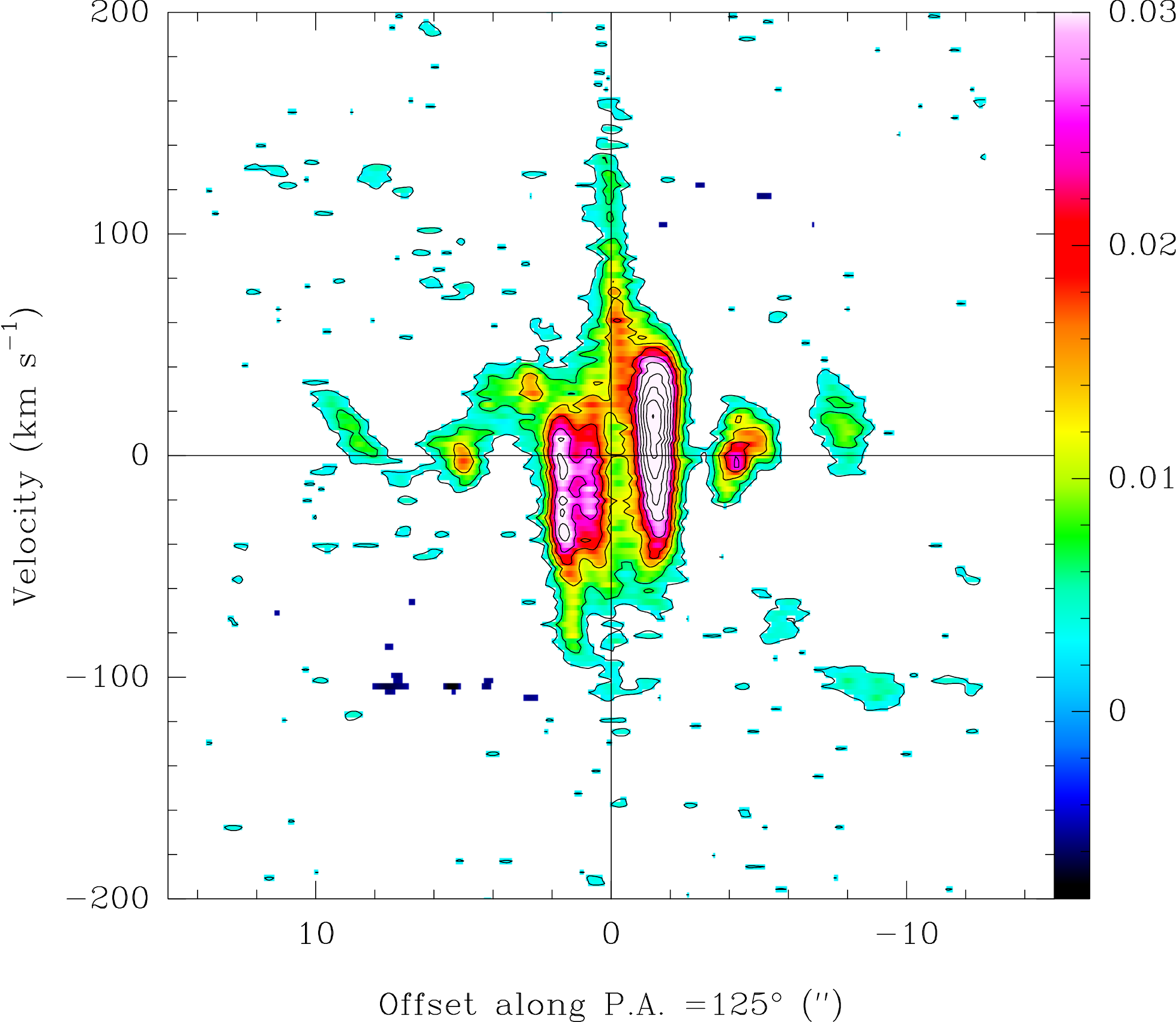}
\caption{Set of position-velocity (PV) diagrams along different PA, selected to capture the different kinematic components seen in Fig. 3. 
The orientation of each slice is marked in Fig. 3 with grey lines. 
Upper panels: PV plot along $\rm PA=35$ deg( i.e. the kinematic major axis) and $\rm PA=40$ deg. 
Lower panels: PV plots along $\rm PA=60$ deg and $\rm PA=125$ deg (i.e. the kinematic minor axis). PA is defined positive starting from East counter-clockwise. The slit width is set to 0.9 arcsec (i.e. approximately the FWHM size of the synthetic beam major axis). The contours are $(2,5,10,15,20,25,30)\sigma$, $\sigma=1.25$ mJy/beam per 2.5 km/s channel. 
}
\end{figure*}

\begin{figure}
\includegraphics[width=\columnwidth]{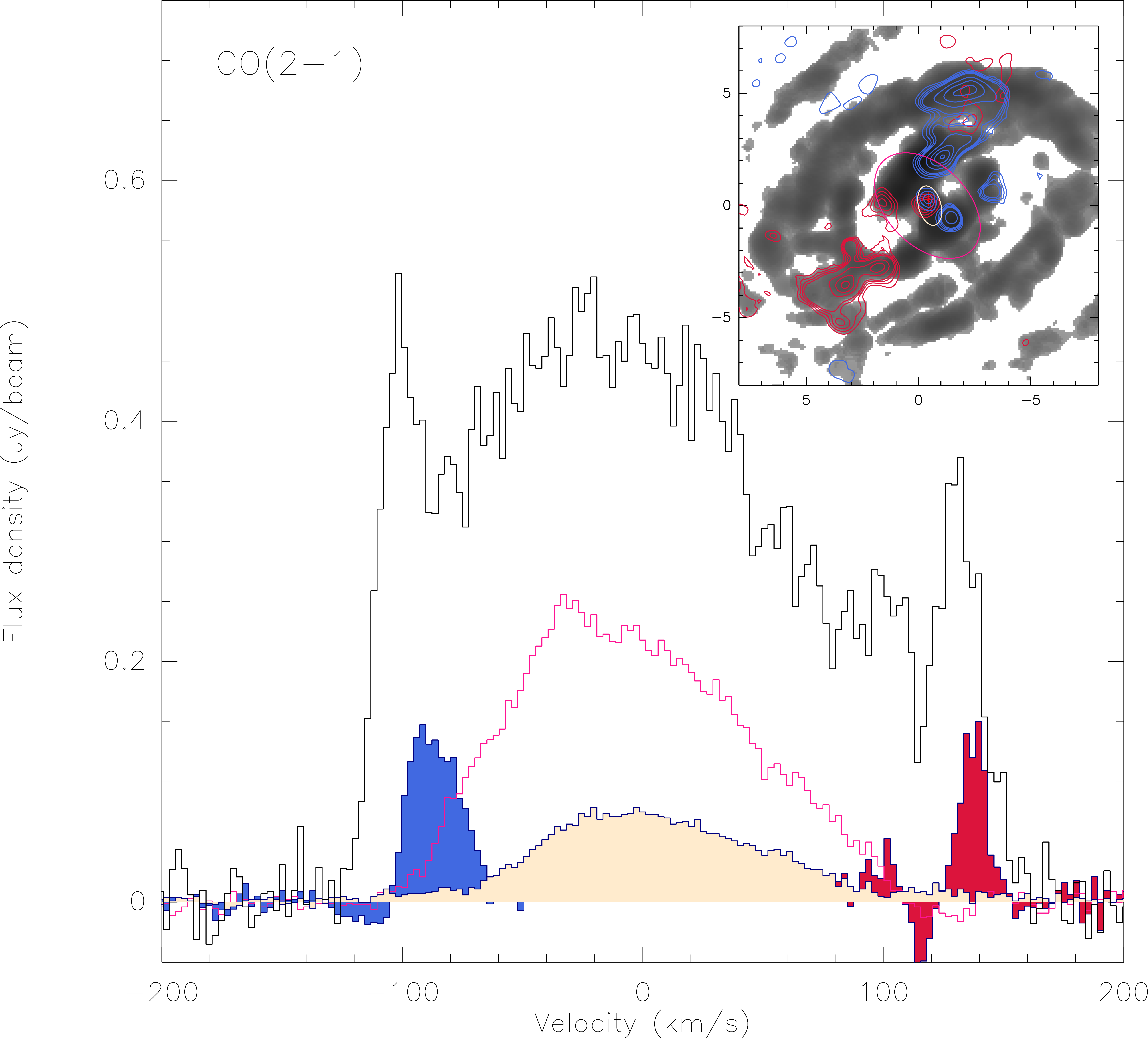}
\caption{The CO(2-1) emission line of ESO428-G14 extracted from different regions. 
Each histogram is colour coded according to the the corresponding region shown in the inset.  
Grey histogram = CO(2-1) spectrum of the galaxy extracted using a mask derived from the $>2\sigma$ region in Fig. 1. 
Deep pink and almond histograms = CO(2-1) extracted from the CNR and nuclear regions, enclosed by the pink and almond ellipses in the inset, respectively. 
Blue and red histograms = CO(2-1) emission extracted from the regions enclosed by the blue and red contours in the inset map. 
Inset panel: the velocity-integrated CO(2-1) emission map (same as Fig. 1). Pink and almond ellipses show the regions from which the respective spectra were extracted. Blue and red contours show the CO(2-1) emitting regions corresponding to the blue and red histograms.  Contours levels are $(2,3,4,5,10,15,20,25,30,25)\sigma$, $\sigma=0.63$ mJy/beam for the blue and $0.8$ mJy/beam for the redshifted components, respectively. 
}
\end{figure}

\begin{figure*}
\centering
\includegraphics[width=14cm]{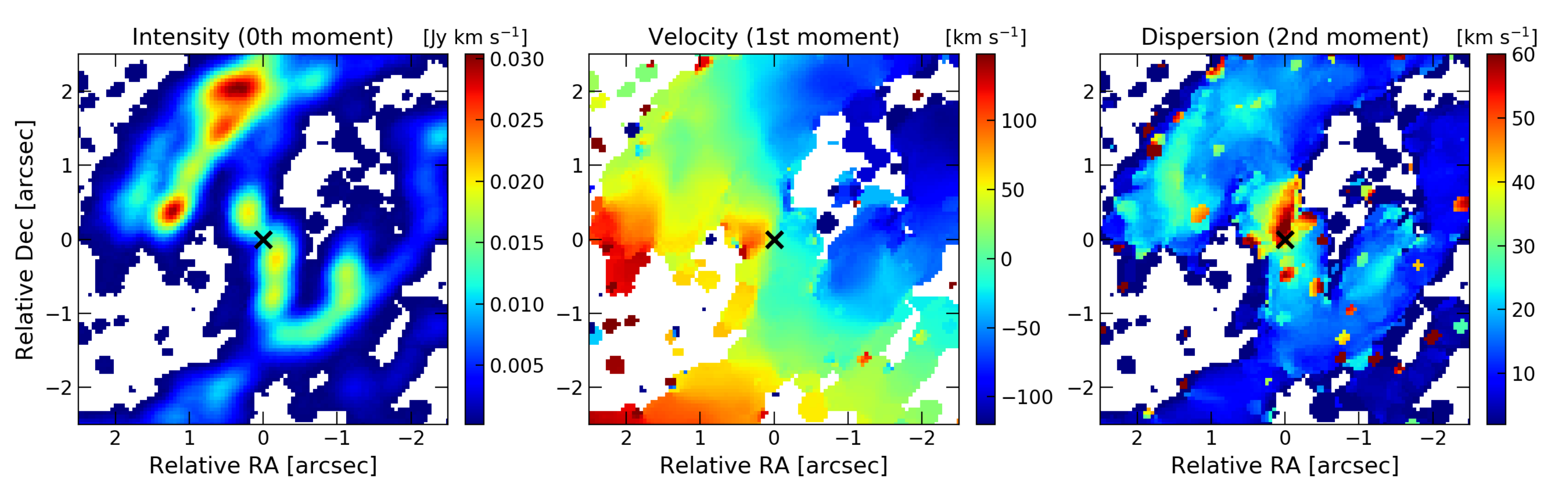}
\includegraphics[width=9cm]{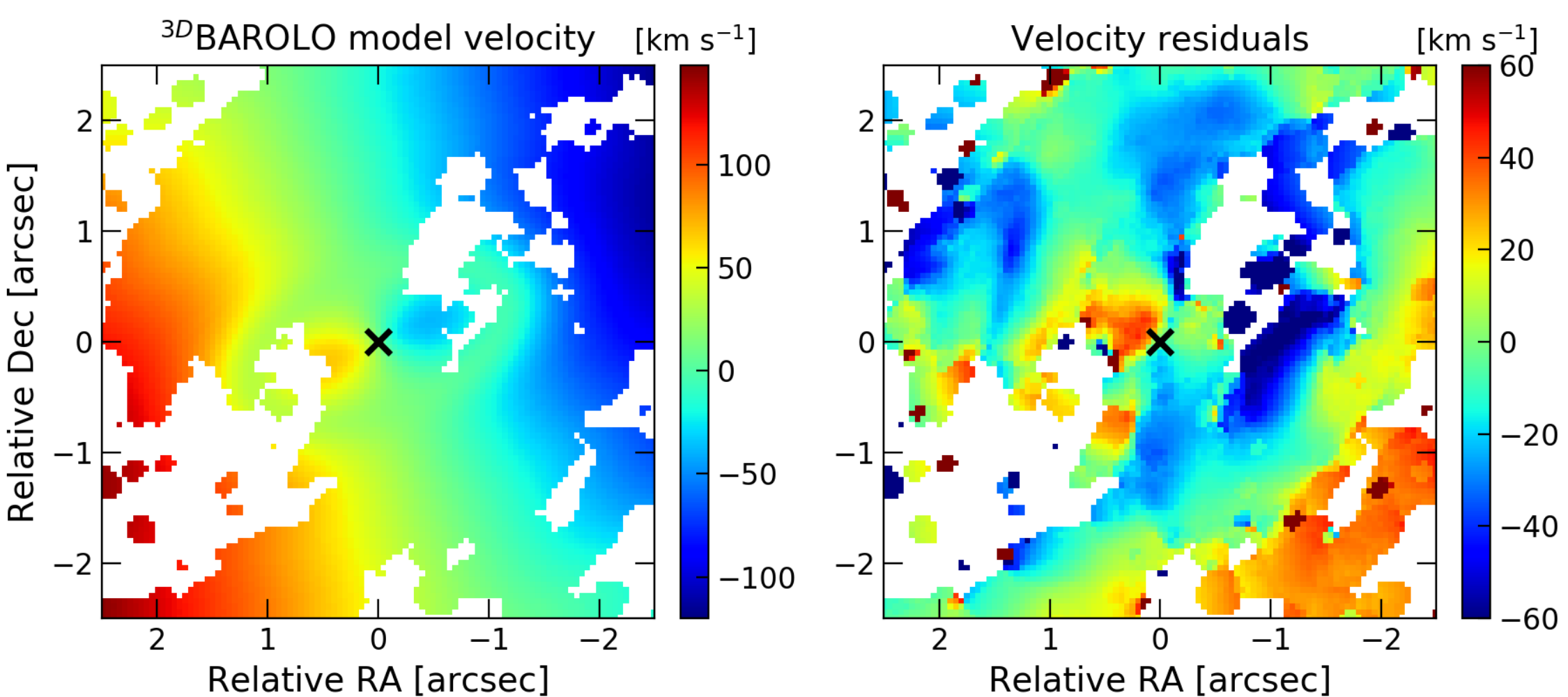}
\caption{Top panels: the CO(2-1) mean-flux (top-left), mean-velocity (top-middle) and velocity dispersion map (top-left), derived for the inner $2.5\times2.5$ arcsec$^2$ region from the high resolution data cube. 
Bottom panels: the $\rm ^{3D}BAROLO$ disk model, and the residual mean-velocity field map. 
The cross indicates the position of the 1.3 mm continuum peak. 
}
\end{figure*}

\subsection{MUSE and SINFONI}

We analysed MUSE@VLT archival data of ESO428-G14. 
The MUSE data were calibrated with the standard pipeline. 
We then used the tool {\it CubeBKGSub} within the CubExtractor software (Borisova et al. 2016) to first estimate and subtract the stellar continuum underlying the emission line of interest, which in this case is [OIII] $\lambda$5007 \AA. 
Then we used {\it CubEx} and {\it Cub2Im} to create a 3D mask and the first moment of the [OIII] flux distribution. 

We also retrieved SINFONI@VLT data from the ESO archive. We used the K-band data described in May et al. (2018) to obtain a map of the $\rm H_2$ 1$-$0 S(1) emission line at 2.12 $\mu$m. 
For the purpose of this work, we limit the analysis to the [OIII] $\lambda$5007 \AA~ and $\rm H_2$ 2.12 $\mu$m emission lines. 
A complete analysis of the MUSE and SINFONI data including several emission lines will be presented in a separate publication.

\section{Results} 

\subsection{Circumnuclear region}

Fig. 1 shows the 1.3 mm continuum and CO(2-1) mean-flux maps. 
The continuum flux density shows a peak at the position of the AGN, and is extended over a region about 5 arcsec wide, located mainly South-East from the nucleus. 
We measure a 1.3 mm continuum flux density of $\rm S_{1.3mm}=3.10\pm0.03$ mJy within the regions above $2\sigma$ (Fig. 1).  

CO(2-1) emission is detected towards the nucleus and in the circumnuclear regions of the host galaxy, out to distances of about a kpc from the AGN. 
The ALMA map reveals clumpy spiral arms that converge to arc-like structures, which may be identified with a circumnuclear ring (CNR) located at a projected radius of about 2 arcsec, and a transverse (equatorial) gas lane which crosses the nucleus and connects the two portions of the CNR. 
Spiral arms and portions thereof are visible at the outskirts of the map. 
The brightest CO(2-1) emission is detected toward a bright clump of gas located about 230 pc north of the AGN (Fig. 1, right panel), which constitutes part of the CNR.  
Continuum emission is also detected at the position of these ark-like structures. 
 
We measure an integrated flux density of $\rm S_{CO}=63.3\pm0.1$ Jy km/s over a line width of 300 km/s, by integrating the flux density within the regions above $2\sigma$ in Fig. 1.  
This translates in to a line luminosity $\rm L^\prime ~CO= 2.1\times10^7~\rm K~km/s~pc^{-2}$. 
By assuming that the CO lines are thermalised, and adopting a luminosity-to-mass conversion factor 
$\rm \alpha_{CO}=3.2~ M_{\odot}~ K^{-1}~(km/s^{-1})~ pc^2$, this translates into a molecular gas mass of 
$\rm M(H_2)=6.7\times10^7~(\alpha_{CO}/3.2)~M_{\odot}$ within the inner $\sim$ 1 kpc radius region.

Fig. 2 shows maps in velocity channels 5 km/s wide, covering the whole CO(3-2) line width. 
The gas with blue-shifted velocity is mainly located at the north-west side of the nucleus, the redshifted one at the south-east side, along a position angle PA $\sim 35$ deg (PA is defined positive from East, increasing counter-clockwise). This defines the kinematic major axis of the galaxy. We detect also deviations from the main velocity gradient at velocity of about $-105\pm10$ km/s on the south-east side, and velocity $129\pm5$ km/s on the north-west side.

Fig. 3 shows the moment 1 and 2 (mean-velocity and velocity dispersion, $\sigma_{disp}$) maps of CO(2-1). 
The data resolve a velocity gradient with a total range of $\sim 250$ km/s from north-west to south-east along a $\rm PA\sim 35$ deg .  
The velocity dispersion $\sigma_{disp}$ is resolved across the galaxy  and is in the range $20-110$ km/s.  
The $\sigma_{disp}$ shows the largest value, exceeding 100 km/s, towards the nucleus, and reaches about 60-80 km/s in the CNR. 
There are also regions of enhanced $\sigma_{disp}$ at large distances from the nucleus, distributed in clumps along $\rm PA=40$ and 60 deg. 
Elsewhere in the outer spiral arms, $\sigma_{disp}=10-30$ km/s. 

We used the $\rm ^{3D}BAROLO$ code (Di Teodoro \& Fraternali 2015) to model the CO(2-1) kinematics of ESO428-G14 with an inclined rotating disk model.  
As first guess parameters for the fit, we input the PA of the observed velocity gradient, $35$ deg, and the inclination of the disk $i=58$ deg, derived from the minor/major axis ratio (NED). 
The model velocity map produced is shown in Fig. 3, bottom-left panel.   
We computed the residual velocity map by subtracting the model from the data (Fig. 3, lower-right panel). 
By inspecting the residual velocity map we find that, while the exponential disk model is able to account for most of the emission seen in the disk and circumnuclear region, the map shows also significant residual red- and blue-shifted emission in a bi-conical structure along PA$\sim60-70$, out to projected distances of 6.5 arcsec from the nucleus (physical scale of about 720 pc).
The best fit disk model has inclination $i=57\pm5$ deg, and circular velocity $v_{rot}=135\pm5$ km/s. 
Fig.  4 shows the rotation velocity $v_{rot}$ versus the radius of the best-fit disk model. 
We find that $v_{rot}$ increases from the centre outwards, shows a peak at radii 2-4 arcsec, and then reaches a plateau at 135 km/s further out in the disk. 

Fig. 5 shows position-velocity diagrams cut along different directions, chosen to capture the main kinematic components seen in Fig. 3.
The PV diagrams are slices taken through the continuum peak position (=AGN position) and using a slit 0.8 arcsec wide, approximately the synthetic beam size (offset along the slit increases from West to East).  
The PV diagrams along the kinematic major axis ($\rm PA=35$ deg) shows the typical disk rotation pattern, quite regular with small deviations from rotating motions, while those with $PA=40, 60$ and 125 deg show increasing kinematic disturbances.
In particular, we detect evidence for radial flows in the inner 1 arcsec with both blueshifted and redshifted projected velocities up to $-120$ kms and 150 km/s, respectively. 
Along $\rm PA=60$ deg we detect non rotational motions with several different velocity at different positions, e.g.  with projected velocity $v=-50$ km/s at offset $-6$ arcsec, with $v=-110$ km/s at offset $+8$ arcsec, and $v=+80$ and $+120$ km/s at offset 6-8 arcsec from the nucleus.

Fig. 6 shows continuum-subtracted spectra of CO(2-1) extracted from different regions on the map. 
The grey histogram is the total CO(2-1) detected in our map, extracted in a region defined by a mask with a threshold of $>2\sigma$ in the velocity-integrated map. The total projected velocity range is from $-$130 to 170 km/s. 
We also show spectra of the CNR (pink histogram), and of the nuclear equatorial gas lane (almond histogram, see caption for details). 
The CNR has  projected velocities in the range $\pm100$ km/s, whereas the gas in the nuclear region reaches velocities exceeding 150 km/s.
We estimate the CO luminosity of the CNR into $\rm L^\prime ~CO=9.6\times10^6 ~\rm K~km/s~pc^{-2}$, and of the nuclear gas lane $\rm L^\prime ~CO =2.8\times10^6~\rm K~km/s~pc^{-2}$, by integrating the total flux density of the respective line (Fig. 6).

We also extract from the residual data cube (i.e. data cube $-$$ \rm ^{3D}BAROLO$ model cube) the CO(2-1) spectra from the regions where we detect the blue- and red-shifted residuals. The spectra are shown in Fig. 6 with blue and red histograms, respectively. 
The corresponding regions are marked in the inset of Fig. 6. In the inset we can see that redshifted and blueshifted emission are detected in a bi-conical like structure out to projected radii of $\sim6.5$ arcsec on both sides of the AGN, as suggested by the mean-velocity residual map (Fig. 3). 
By fitting the spectra with Gaussian functions, we find that the red peaks at a projected velocity of $v_{red}=137\pm1$ km/s with $\rm FWHM=15$ km/s, and the blue at $v_{blue}=-87\pm0.2$ km/s with $\rm FWHM=24$ km/s. 
We derive integrated flux densities of  $Sdv_{red}=2$ Jy km/s and $Sdv_{blue}=3.8$ Jy km/s. 
These correspond to line luminosities of $\rm L^\prime ~CO_{red} =(6.6 \pm 0.5)\times 10^5~\rm K~km/s~pc^{-2}$ and $\rm L^\prime ~CO_{blue} =(1.3\pm0.1)\times10^6~\rm K~km/s~pc^{-2}$.

\begin{figure}[t]
\includegraphics[width=8cm]{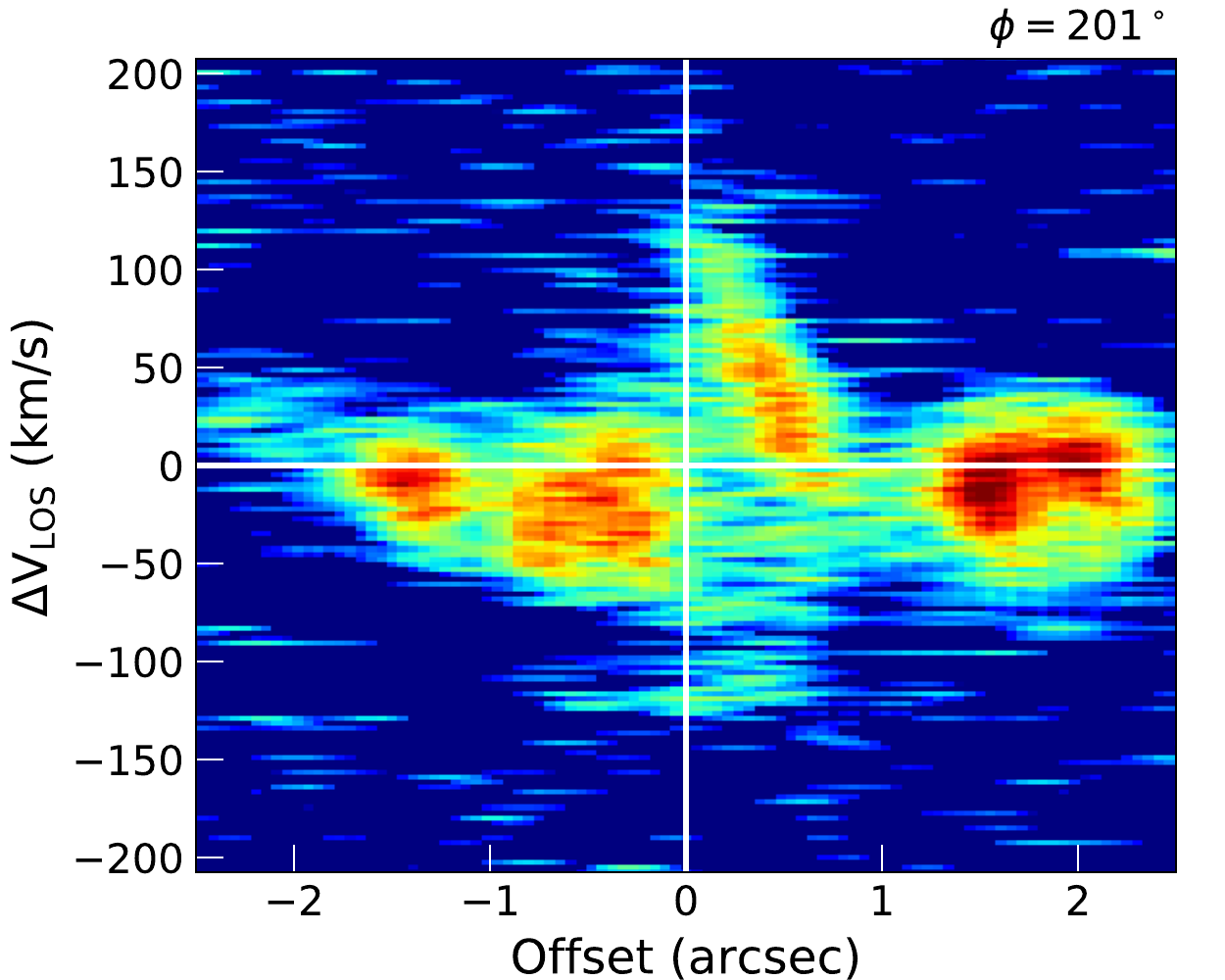}
\caption{The PV diagram of the nuclear region along the kinematic minor axis. Deviations from the rotation velocity are detected in the inner 1 arcsec, where molecular gas with velocity up to $150$ and $-130$ km/s is detected. 
}
\end{figure}

\begin{figure*}[t]
\centering
\includegraphics[width=14cm]{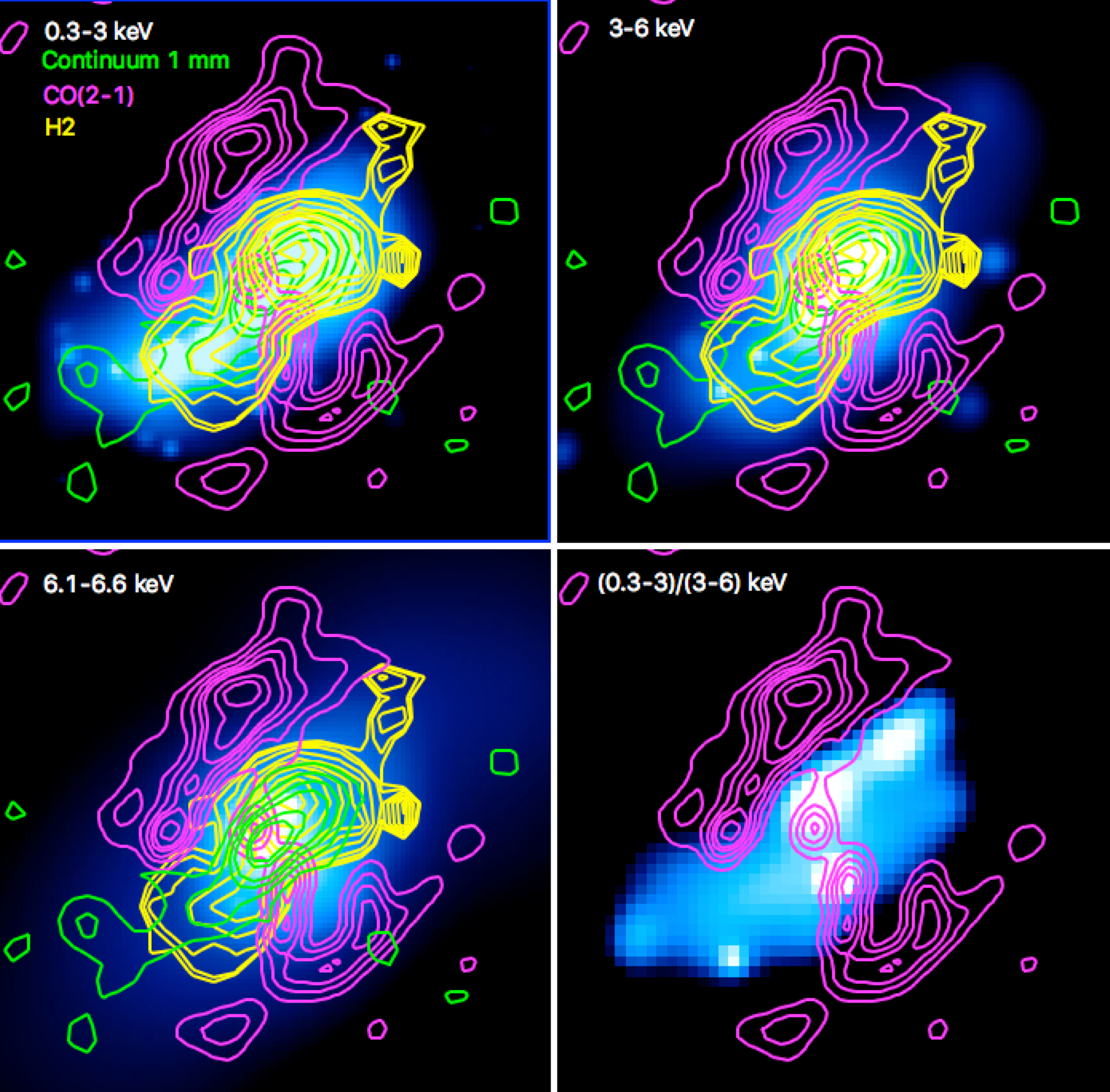}
\caption{Zoom-in towards the nuclear region of ESO428-G14. Color images are the Chandra soft (upper-left panel), hard band (upper-right panel), and Fe K$\alpha$ line emission (lower-left panel), and 
hardness ratio maps (from Fabbiano et al. 2018). Green contours = 1.3 mm continuum detected by ALMA imaged with robust weighting. Magenta contours =  velocity-integrated CO(2-1) line emission imaged with robust weighting to enhance the angular resolution in the nuclear region. Yellow contours = warm molecular gas traced by the H$_2$ 2.12 $\mu$m line from SINFONI/VLT data. 
}

\end{figure*}

\subsection{Nuclear region}

To better investigate the morphology and kinematics of the inner 1 arcsec ($\sim 100$ pc) region, we used the data cube with enhanced angular resolution (Section 2). 
Fig. 7 (upper panels) shows the mean-flux, mean-velocity and velocity dispersion maps within the inner $5\times5$ arcsec region around the AGN. 
Arc-like CO-emitting structures and gas lanes that come in from the circumnuclear ring towards the AGN are detected. 
The inner CO(2-1) gas lane is resolved into three main clumps.   
We modelled the emission in this region with a inclined exponential disk with fixed inclination and PA as in the main outer disk. 
Figure 7 (lower panels) show the disk model and the $data-model$ residual velocity map.  
High resolution data confirm the positive residual toward the nucleus, which is not consistent with rotation of the disk in the inner region. 
This is better seen in the PV diagram extracted along the kinematic minor axis (Fig. 8), where molecular gas with velocity up to $-130$ and $150$ km/s is detected in the inner $\sim100$ pc.  
In this inner region the molecular gas kinematics deviates from the rotation pattern. 
The PV diagram suggests that the gas comes in from the northern arm of the CNR with velocity close to the rotation velocity of the CNR, and its velocity progressively increases while approaching the AGN (located at 0 offset).

We compare the distributions of the cold gas and dust emission seen in the ALMA data in the nuclear region, with that of the X-ray and $\rm H_2$ emitting material detected with Chandra and SINFONI/VLT. 
Fig. 9 shows colour images in Chandra energy bands, over which contours of the 1.3 mm continuum, CO(2-1) and $\rm H_2$ emission have been overplotted. 
The 1.3 mm continuum emission (green contours) peaks close to the X-ray nucleus and partly overlaps with the extended soft X-ray emission towards south-east.
Both the X-ray and 1.3 mm continuum in this region are distributed along  a PA similar to that of  the ionisation cone and radio jet (Ulvestad \& Wilson 1989, Falcke, Wilson \& Simpson 1998), and 2.12 $\mu$m H$_2$-emitting gas (yellow contours, May et al. 2018).  

The inner transverse CO gas lane overlaps with the most obscured, Compton-thick region seen in X-rays (Fig. 7 in Fabbiano et al. 2018a, and Fig. 9, lower-right panel). 
The CO emission from the CNR is not co-spatial with the X-ray emission, which instead extends also beyond the CO-emitting gas.  
This configuration appears to form a CO-cavity, which is filled with X-ray emitting material and with $\rm H_2$ gas.

\begin{figure*}[t]
\centering
\includegraphics[width=\columnwidth]{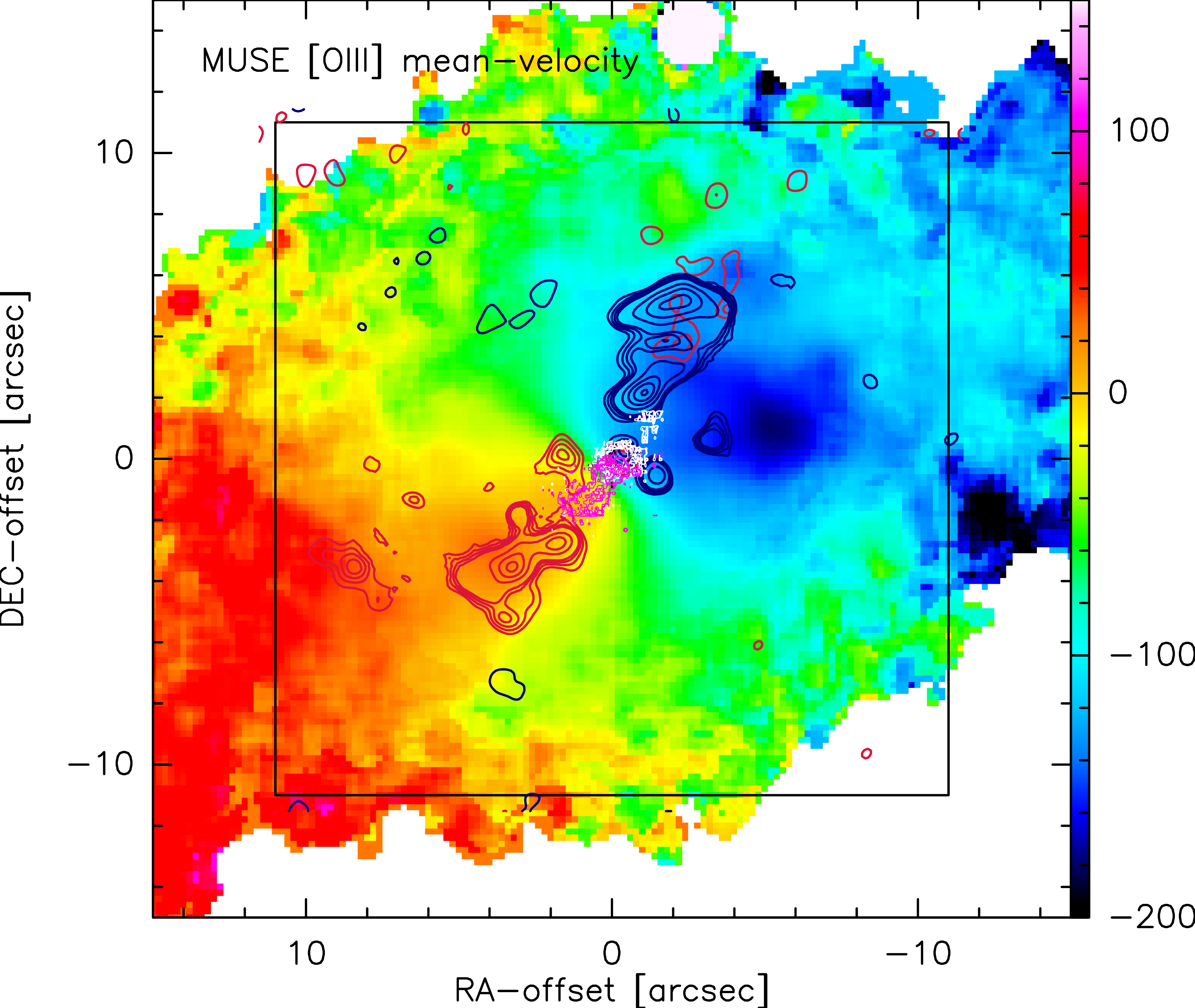}
\includegraphics[width=\columnwidth]{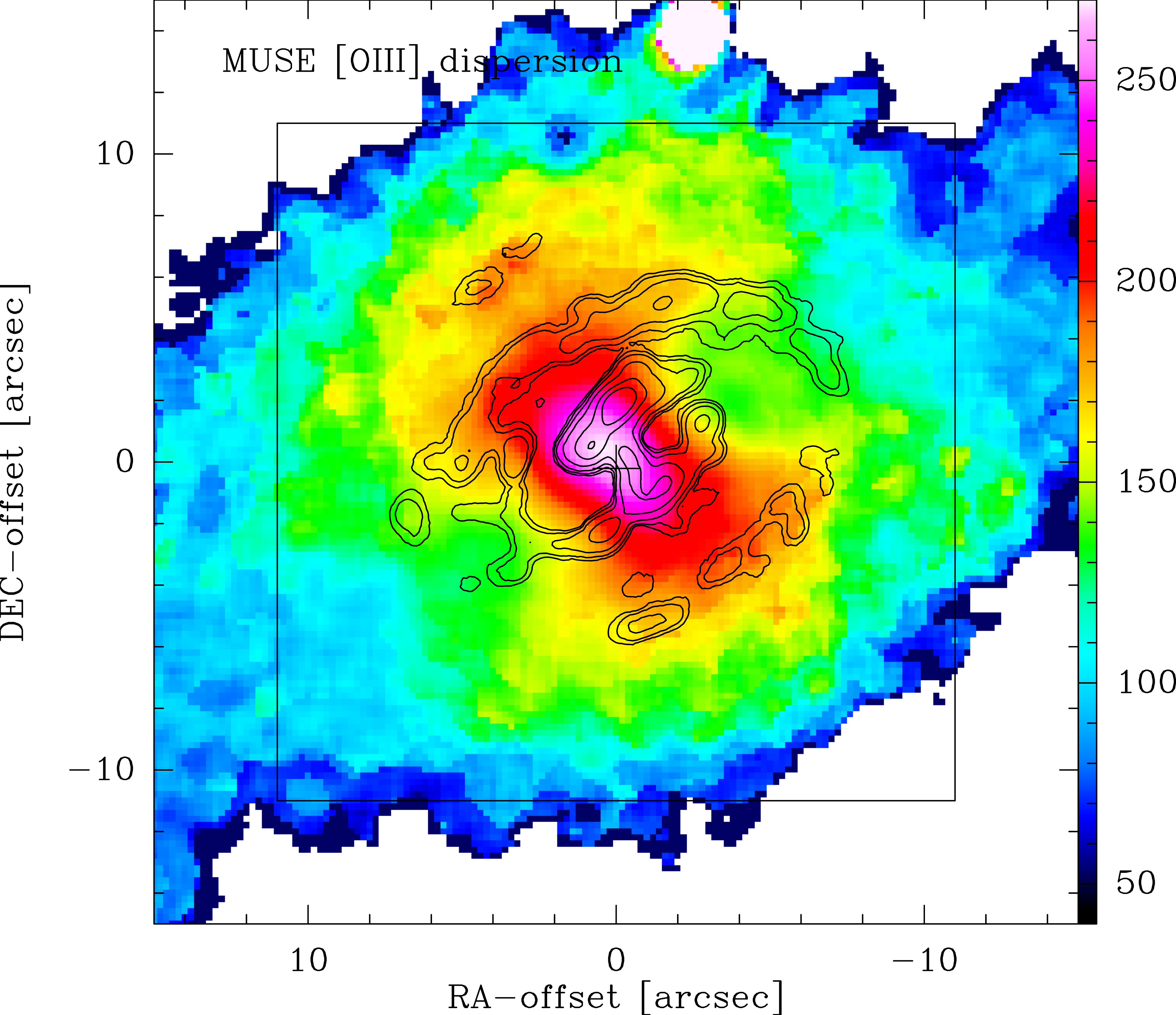}
\caption{Left panel: [OIII]$\lambda$5007 $\AA$~ mean-velocity map of ESO428-G14 obtained with MUSE/VLT.  
Red/blue contours = receding/approaching CO outflow. Contours are as in Fig. 6.
White/magenta contours= receding/approaching sides of the $\rm H_2$ emission line from SINFONI data. 
The black box shows the $22\times 22$ arcsec field of view show in Fig. 3. 
Right panel: the [OIII] velocity dispersion map. Overplot with black contours is the CO(2-1) mean-flux map. 
Color scale units are km/s. 
}
\end{figure*}

\section{Discussion}

\subsection{Molecular disk}

We measure a molecular gas reservoir of $\rm M(H_2)=6.7\times 10^7 (\alpha_{CO}/3.2)~M_{\odot}$. However it is difficult to exactly quantify the fraction of flux which may have been filtered out by the interferometer on large scales. 
The brightest CO emission is detected in a circumnuclear ring (CNR) with radius 200 pc and mass $\rm M(H_2)=3\times10^7 (\alpha_{CO}/3.2)~M_{\odot}$.
We interpret this structure as the inner Lindblad resonance region, in which, according to Combes \& Gerin (1985), the gas can get stalled in ring-like structures, before being fed towards the nucleus.

Regarding the kinematics of the molecular gas in the host galaxy, 
most of the observed velocity gradient on large scales is well modelled with a exponential disk seen at an inclination of $i=57$ deg, and with $v_{rot}=135$ km/s. 
Based on our modelling, we derive a dynamical mass $M_{dyn }=R \times v_{rot}^2/G=5\times 10^9~\rm M_{\odot}$ within a radius of $\sim 1$ kpc.

We find that the $\sigma_{disp}/v_{rot}$ varies between $\sim 0.1$ in the spiral arms, $0.8-1$ in the CNR, and exceeds $3$ in the inner $\sim 100$ pc region, indicating that the gas is turbulent in the circumnuclear and nuclear regions, and subject to inflow (as we will discuss in Section 4.2).

We show in Fig. 10 the [OIII]$\lambda$5007 \AA~ velocity and velocity dispersion map from MUSE, which traces the distribution and kinematics of ionised gas in the host galaxy. 
[OIII] shows a mean-velocity gradient  with similar PA to that observed for CO. 
We note that the MUSE FOV is larger than the ALMA primary beam of these observations, and that the [OIII] emission extends further out and has a smoother distribution (no spiral arms or rings are detected in [OIII]). 

The ionised gas traced by [OIII] shows much larger velocity dispersion than the molecular gas, reaching $\sigma_{disp,[OIII]}\sim180-200$ km/s  at the location of the outer CO arm, and exceeding 200 km/s in the CNR. 
 Interestingly, the regions with the largest [OIII] velocity dispersion trace well the molecular spiral arms, CNR and inner gas lane.

\subsection{Feeding}

In the nuclear region, the two brightest portions of the CNR appear connected by a transverse gas lane (or bar), which goes through the nucleus. 
We estimate a mass of molecular gas in the inner bar of $\rm M(H_2)=9\times 10^6 (\alpha_{CO}/3.2)~ M_{\odot}$ within a region of radius $\sim50$ pc. 
In this inner region, we detect gas with projected velocity out to $150$ km/s and $-130$ km/s (Fig. 4 and 8). 
The CO kinematics in this region deviates from rotation, and shows the typical pattern due to radial inflow or outflow.  
The PV plot along the kinematic minor axis (Fig. 5) is similar to those found in other nearby galaxies like, e.g.,  NGC1566 and NGC1961 (Combes et al. 2009, 2014), NGC 4826 (Garcia-Burillo et al. 2003), NGC3147 (Casasola et al. 2008), NGC3393 (Finlez et al. 2018).
Fig. 8 shows that the nuclear fast gas has maximum projected velocity at the AGN location, and its velocity decreases towards the location of the CNR. 
This configuration suggests that the gas comes in from the CNR, and forms an inflow of material towards the AGN. 

 We compute the inflow rate within a region of $R_{in}=100$ pc as follows. 
The inflow rate is $\rm \dot M_{in}=v_{max,R_{in}}\times M(H_{2})/R_{in}= 2 ~  (\alpha_{CO}/3.2)~M_{\odot}/yr $, where $R_{in}=100$ pc and $v_{max,R_{in}}=50$ km/s, defined by fitting a Gaussian function to the nuclear spectrum in Fig. 6 (almond histogram). 
The Bondi radius of the BH in ESO428-G14 is of the order of 1 pc, much smaller than the size of the inner gas lane, therefore the molecular gas can undergo significant fragmentation and star formation before reaching the nucleus, if the dynamical timescale is longer than the star formation depletion timescale, $\rm t_{dep}=M_{gas}/SFR$. 
We evaluated the SFR in the inner 100 pc from the 1.3 mm continuum. Adopting the M82 spectral energy distribution, we compute the far-infrared luminosity of ESO428-G14 and convert it into a  $\rm SFR=0.3 ~M_{\odot}/yr$. 
We also measure the narrow  $\rm H\alpha$
luminosity in this central region from the MUSE data, which we convert into a SFR using the relation $\rm SFR(H\alpha)(M_{\odot}/yr)=7.9\times10^{-42}~L_{H\alpha} ~(erg/s)$ (Kennicutt et al. 1994). 
From this relation we derive $\rm SFR_{H\alpha}=0.02~M_{\odot}/yr$. 
$\rm SFR_{H\alpha}$ is probably affected by extinction, therefore we adopt the $\rm SFR=0.3 ~M_{\odot}/yr$ as our fiducial value. 
We derive therefore a gas depletion timescale $\rm t_{dep}=M_{gas}/SFR=30~Myr$. 
The dynamical timescale of the gas within 100 pc radius is $\rm t_{dyn}=R/v(R)\approx2~Myr$, where $\rm v(R)=50~ km/s$ is the maximum velocity of the gas at R=100 pc (Fig. 8). 
We find that $t_{dep}<<t_{dyn}$,  suggesting that most of the inflowing molecular gas in the inner bar does not fuel nuclear star formation.

The BH accretion rate inferred from the bolometric luminosity $\rm L_{bol}$,  $\rm \dot M_{BH}=L_{bol}/\epsilon c^2=0.007~M_{\odot}/yr$ for a radiatively efficient accretion ($\epsilon=0.1$), or $0.07~M_{\odot}/yr$ for a radiatively inefficient accretion ($\epsilon=0.01$). Even for the latter case, $\rm \dot M_{BH}<\dot M_{in}$, suggesting that a large fraction of the gas within $R_{in}$ does not reach the black hole. 
A possible solution is that a large fraction of the inflowing gas is lost in outflows (see Section 4.3), thus reducing the net accretion rate reaching the BH,  $\dot M_{BH}$, as predicted for both radiatively efficient accretion flows (e.g Kurosawa \& Proga 2009) and radiatively inefficient flows (e.g. Sadowski et al. 2016, Yuan et al. 2015).


Interestingly, the inner CO bar overlaps with the most obscured, Compton-thick region seen in X-rays (Fig. 7 in Fabbiano et al. 2018a, and Fig. 9, lower-right panel). 
We estimate the line-of-sight $\rm H_2$ column density of $\rm N(H_2) \approx 2\times10^{23}~ cm^{-2}$ toward the the inner bar region, which has approximate size of 100 by 50 pc$^2$. 
This is similar to the $\rm N(H_2)$ found in other local Seyfert galaxies (Alonso-Herrero et al. 2018, Izumi et al. 2018), and smaller than the neutral hydrogen column density $\rm N_H$ towards the AGN, as derived from X-ray observations, which is $>10^{24}~\rm cm^{-2}$ (Fabbiano et al. 2017, 2018b).  This $\rm N(H_2)$ is a lower limit of the true column density due to beam dilution.
In addition, $\rm N(H_2)$ is derived for a given conversion factor ($\alpha_{CO}=3.2~ \rm M_{\odot}~ K^{-1}~(km/s^{-1})~ pc^2$). 
While there are empirical evidences that in the nuclei of galaxies conversion factor is smaller than the Milky Way (MW) value and can be of the order 1, there have been also claims that it can be a factor of a few higher then in the MW (e.g. Wada et al. 2018). 
The uncertainties on $\rm M(H_2)$, and consequently on $\rm N(H_2)$, are therefore of the order of a few. 
It is thus possible that the molecular nuclear gas contributes significantly to the Compton thickness of this AGN. 

The nuclear equatorial lane may be identified with a inner bar, or with the (outer portion of the) molecular torus, similarly to what found in in NGC5643 by Fabbiano et al. (2018c) and  Alonso-Herrero et al. (2018). 
In that AGN, the nuclear molecular disk also matches the regions with the largest nuclear obscuration seen in X-rays. 
Higher angular resolution observations are required to confirm this scenario and map in detail the distribution of the molecular emission in the nucleus of ESO428-G14.

\subsection{Feedback}

The nuclear configuration of the CO and X-ray emitting gas in ESO428-G14 resembles the case of NGC1068 as studied by Garcia-Burillo et al. (2010).  
They found that hard X-ray emission overlaps with the molecular circum-nuclear disk, but also extends further out from it, like in the case of ESO428-G14.
These CO cavities are puzzling because the hard (3-6 keV) continuum and Fe K$\alpha$ emission require scattering from dense neutral clouds in the ISM (Fabbiano et al. 2018), therefore we would expect to detect CO in the region detected in hard X-ray in ESO428-G14. 
The upper limit derived from these observations implies that the mass of molecular gas in such dense clouds must be smaller than $\rm M(H_2)<7\times10^5~ (\alpha_{CO}/3.2)~ M_{\odot}$. 
It is possible that in this region CO(2-1) emission is suppressed due to strong AGN irradiation and/or shocks. 
The detection of warm molecular gas traced by the $\rm H_2$ 2.12 $\mu m$ emission line in the CO cavity supports this scenario, confirming that the scattering material giving rise to the hard X-ray extended emission is warm molecular gas, as suggested by Fabbiano et a. (2018). 
A similar case of warm $\rm H_2$-emitting gas filling a CO-deprived region, has been recently reported in the AGN host galaxy NGC2110 (Rosario et al. 2019). 
In ESO428-G14, CO may be excited to higher excitation levels up the CO ladder in this region, due to irradiation by hard X-ray field and/or shocks.

At larger distances from the nucleus, several kinematic disturbances are detected in the mean-velocity and velocity dispersion maps, which indicate outflowing gas. 
The PV diagrams along PA =40-60 deg (Fig. 5) show that several clumps of non-rotating molecular material are detected at projected distances of  $\sim2$ to 7 arcsec  (i.e. 0.2-0.7 kpc) from the AGN with projected velocities out to $+130$ km/s and  $-110$ km/s. 
Figure 2, 3 and 6 show that these clumps are distributed in a bi-conical pattern. 
Because the PA of the outflow is similar to that of the main velocity gradient of the disk and spiral arms, in the outer parts the outflow overlaps with the rotation pattern in both space and velocity, making it challenging to separate the two kinematic components. 
We interpret these features as the two sections of a bi-conical outflow.   
The projected distance reached by the outflow is about 700 pc on in both sides of the AGN. 
The mass of molecular gas in the two cones is estimated into $\rm M(H_2)_{blue}=5.3\times10^5~(\alpha_{CO}/0.8)~M_{\odot}$ and $\rm M(H_2)_{red}=10^6~(\alpha_{CO}/0.8)~M_{\odot}$.   
 
We derive the outflow rate through the relation 
\begin{equation}
\dot M_{of}=3 \frac{M_{of}~v_{max}}{R_{of}}
\end{equation}
(Fiore et al. 2017), where the maximum outflow velocity  $\rm v_{max}=v+2\sigma$ is estimated from the Gaussian fit of the lines (Fig. 6).
We find a molecular outflow rate of  $\rm \dot M_{of}\approx0.8~(\alpha_{CO}/0.8)~M_{\odot}/yr$, considering the global contribution of both sides of the bi-cone.

An outflow on similar scales is also seen in ionised gas traced by [OIII] (Falcke et al. 1996).
Riffel et al. (2007) suggested that the radio jet is launched at a small angle relative to the galactic plane, thus perturbing the kinematics of the ionised gas in the disk. 
In Fig. 10 we show the MUSE [OIII] velocity map with the contours of the CO outflow emission bi-cones overlayed. 
We also overplot the velocity-integrated red- and blue- sides of the $\rm H_2$ 2.12 $\mu m$ line derived from SINFONI data (white and magenta contours). 
May et al. (2018) detected a nuclear outflow using the [Si VI] $\lambda19641~ \AA$ coronal line from the same SINFONI data set, and they estimated a outflow rate of warm molecular gas of 3-8 $\rm M_{\odot}/yr$ within a region of 170 pc radius. 
We show in Fig. 10 that the warm $\rm H_2$ outflow traces the inner portion of the bi-conical outflow detected in CO, suggesting that the outflow cools with increasing distance from the AGN. 
Both the CO and the $\rm H_2$ outflows appear slightly tilted with respect to the kinematic major axis, and are approximately aligned with the [OIII] ionisation cone direction. 
A complete analysis of the [OIII] kinematics, separating the contribution of the host galaxy rotating disk from the AGN ionisation cones and outflows, will be presented in a separate paper.

\section{Conclusions}

We imaged the molecular gas of the host galaxy of the nearby Compton-thick AGN ESO428-G14 at 70 pc resolution using ALMA archival data of CO(2-1) line and 230 GHz continuum. We joined ALMA data with Chandra X-ray data, MUSE/VLT [OIII] and $H_2$ maps from SINFONI/VLT, and derived the following conclusions. 

\begin{itemize}

\item
The molecular gas in the host galaxy is distributed in rotating disk with inclined at  $i=57$ deg, and with 
$\rm v_{rot}=135$ km/s. We derive a dynamical mass $\rm M_{dyn }=5\times 10^9~\rm M_{\odot}$ within a radius of $\sim 1$ kpc.

\item The CO emission is distributed in clumpy spiral arms out to radii of about 1 kpc. 
The brightest CO emission is detected in a circumnuclear ring (CNR) with radius $\approx 200$ pc and mass $\rm M(H_2)=3\times10^7 (\alpha_{CO}/3.2)~M_{\odot}$.
We interpret this structure as the inner Lindblad resonance region, in which the gas can get stalled in ring-like structures, before being fed towards the nucleus.

\item
In the inner 100 pc region the observed kinematics is consistent with a nuclear inflow, or inner bar, which feeds the AGN. 
The inner CO bar overlaps with the most obscured, Compton-thick region seen in X-rays, and may contribute significantly to the Compton thickness of this AGN. 
Observations with higher angular resolution are required to map in detail the kinematics in this innermost region around the AGN. 

\item We detect a molecular outflow with a outflow rate of  $\rm \dot M_{of}\approx 0.8~(\alpha_{CO}/0.8)~M_{\odot}/yr$, distributed along a bi-conical structure  with projected size of $700$ pc on both sides of the AGN, approximately along the direction of the [OIII] ionisation cones, and along the same direction of the $\rm H_2$ outflow reported by May et al. (2018). The warm molecular outflow traced by $\rm H_2$ is detected in the inner 170 pc, whereas the CO outflow is detected out to 700 pc, indicating that the gas cools down while leaving the nucleus.

\item
The hard band 3-6 keV X-ray emission overlaps with the molecular circumnuclear ring detected by ALMA, but also extends further out where no CO-emitting gas is detected. 
The hard X-ray emitting CO cavity is filled with warm molecular gas, traced by the  2.12 $\mu$m $\rm H_2$ emission line. 
This confirms that the hard (3-6 keV) continuum and Fe K$\alpha$ emission are due to scattering from dense neutral clouds in the ISM. 
This region hence contains molecular material, but may appear CO-deprived due to irradiation by the AGN hard X-ray field and/or shocks, which may excite CO to higher levels up the rotational ladder. 
Observations of higher J CO transitions may be useful to map the molecular gas and constrain the effects of AGN feeding \& feedback in the inner nuclear region.

\end{itemize}

\begin{acknowledgements}
This paper makes use of the following ALMA data: ADS/JAO.ALMA\#2015.1.00086.S. ALMA is a partnership of ESO (representing its member states), NSF (USA) and NINS (Japan), together with NRC (Canada), MOST and ASIAA (Taiwan), and KASI (Republic of Korea), in cooperation with the Republic of Chile. The Joint ALMA Observatory is operated by ESO, AUI/NRAO and NAOJ.   
Based on observations collected at the European Southern Observatory under ESO programmes 086.B-0484(A) and 097.D-0408.
This work was partially supported by the Chandra Guest Observer program grant GO5-16090X (PI: Fabbiano).
CF, FF, CF, MB, AT acknowledge support from INAF PRIN SKA Forecast and ASI INAF contract  I/037/12/1-2016. We thank Angela Malizia and Alessandro Marconi for insightful discussion.
 \end{acknowledgements}

\end{document}